\documentclass[letterpaper,twocolumn,10pt]{article}
\usepackage{usenix2019_v3}

\usepackage{tikz}
\usepackage{amsmath}
\usepackage{caption}
\usepackage{subcaption}
\usepackage{graphicx}
\usepackage{diagbox}
\usepackage{listings}
\usepackage{tcolorbox}
\usepackage{enumitem}
\usepackage{xspace}

\hypersetup{draft}

\definecolor{mygreen}{rgb}{0,0.6,0}
\definecolor{mygray}{rgb}{0.5,0.5,0.5}
\definecolor{mymauve}{rgb}{0.58,0,0.82}

\lstdefinestyle{top}{
  float=tp,
  floatplacement=tbp
}

\newcommand{\name}{SDNator\xspace}
\newcommand{\etc}{\textit{etc.}\xspace}
\newcommand{\ie}{\textit{i.e.,}\xspace}
\newcommand{\eg}{\textit{e.g.,}\xspace}
\newcommand{\etal}{\textit{et al.}\xspace}

\newcommand{\aka}{\textit{a.k.a.}\xspace}

\usepackage{etoolbox}
\makeatletter
\def\@maketitle{\newpage
 \vbox to 3.0in{
 \vskip 0.2in
 \begin{center}%
  {\Large\bf \@title \par}%
  \vskip 0.28in
  {\large\it
   \lineskip .5em
   \begin{tabular}[t]{c}\@author
   \end{tabular}\par}%
 \end{center}%
 \par
 \vfil
 }
}
\makeatother

\begin{document}

\date{}


\title{\Large \bf SDNator is Not Another SDN Controller:\\
Enabling Extensible Data-Driven Control in Cyber-Physical Systems}




\author{
{\rm Yikai Lin}\\
University of Michigan
\and
{\rm Ruowang Zhang}\\
University of Michigan
\and
{\rm Efe Balta}\\
University of Michigan
\and
{\rm Jianbin Zhang}\\
University of Michigan
\and
{\rm Xiao Zhu}\\
University of Michigan
\and
{\rm Kira Barton}\\
University of Michigan
\and
{\rm Dawn Tilbury}\\
University of Michigan
\and
{\rm Z. Morley Mao}\\
University of Michigan
}

\maketitle

\begin{abstract}
An SDN-like centralized control architecture is increasingly popular and has been widely explored in cyber-physical systems (CPS) such as manufacturing, internet-of-things, and autonomous vehicle systems for higher flexibility, programmability and scalability. However, no existing frameworks can offer  domain-agnostic, easily extensible support for data-driven CPS applications. In this work, we design, implement, and open-source \textit{SDNator}, the first framework to enable extensible, data-driven control in CPS. SDNator embraces an application- and data-driven design where applications function as data consumers and producers to collectively define the workflows of the controller. SDNator also incorporates two data store backends to support both event-driven and data-driven programming patterns. Benchmarks show that SDNator is highly scalable, and delivers comparable performance to Ryu, a widely used SDN controller.\looseness=-1

Moreover, we demonstrate the capabilities and usability of SDNator through our case studies of manufacturing and networking systems. By integrating applications from respective domains, we build different ``controllers'' for different scenarios. Most notably, we leverage SDNator to implement the first digital-twin-equipped central controller for additive manufacturing fleets. We show through extensive and realistic simulations that SDNator-based scheduling can (1) significantly shorten production time and improve reliability in the presence of anomalies compared to decentralized approaches, and (2) flexibly adjust and optimize production plans upon urgent requests such as producing Personal Protective Equipment during the COVID-19 pandemic.
\end{abstract}

\vspace{-3mm}

\section{Introduction}
\label{sec:intro}

\vspace{-2mm}

Recent advances in hardware capabilities (\eg programmable switches, autonomous automobile, smart manufacturing) and network connectivity technologies (\eg 4G and 5G) have enabled new software-hardware interactions in cyber-physical systems (CPS). For instance, a centralized control architecture inspired by Software-Defined Networks (SDN) has been widely explored as a viable alternative to traditional distributed solutions in various CPS such as manufacturing\cite{lopez2018sdc,qamsane2019unified,balta2018centralized}, internet-of-things (IoT)\cite{liu2015sdiot}, and autonomous driving\cite{jindal2018sedative,liu2016vanet}. Applications can utilize aggregated real-time information of a CPS to improve its flexibility, efficiency and reliability. Furthermore, this paradigm opens up exciting research opportunities for new control strategies and workflows that were previously infeasible.

However, developing and deploying CPS applications based on this centralized paradigm is non-trivial. Due to a lack of available ``controller'' frameworks that readily support CPS applications, existing works often resort to extension or adaptation of legacy SDN controllers to tailor to their particular application requirements\cite{lopez2018sdc,liu2015sdiot,kathiravelu2019sd}. Such extension or adaptation is time-consuming, non-reusable, and inherits the same limitation in supporting data-driven CPS  applications~(\ie~inefficient data production/consumption, lack of data heterogeneity and historical data support,~\etc\cite{lin2019add}). If a new ``controller'' variant has to be built every time new CPS applications emerge, which is likely given how heterogeneous CPS are, the value of flexibility and programmability diminishes.\looseness=-1



To fundamentally address this problem, we design, implement and open-source \name(\S\ref{sec:design}), a framework that enables researchers and developers to easily implement new or incorporate existing CPS applications in a centralized workflow. SDNator is not a specialized controller; rather, it enables building different controllers through plug-and-play of different applications/systems. \name embraces an application- and data-driven design where applications function as data consumers and producers to collectively define the workflows and capabilities of a controller. A data-driven design also allows \name to be domain- and protocol-independent, making it possible to integrate with and be deployed in real-world CPS. 

Enlightened by previous work on CPS application characteristics and requirements\cite{lin2019add}, \name incorporates two different data backends to provide both event-driven and data-driven programming patterns. One of them serves as persistent storage to enable historical data usage for offline analysis. To make \name more instrumental and efficient, we implement an on-demand data production (\S\ref{sec:demand}) mechanism to give consumer apps control over what data will be produced at what frequency. Fault tolerance and recovery features (\S\ref{sec:fault}) are also built into \name for extra reliability. Through extensive benchmarks~(\S\ref{sec:benchmark}), we show that \name (1) delivers over 100K msgs/s using one CPU core on a commodity PC, (2) incurs as little as sub-100us end-to-end delay, and (3) scales with 60 geo-distributed apps and high network latency.

On top of that, we leverage \name to implement 3 different control workflows for manufacturing (\S\ref{sec:manufacturing}) and networking (\S\ref{sec:networking}) CPS in our case studies. In particular, in light of the ``citizen-supply-chain'' phenomenon\cite{url:citizen-ppe} during the COVID-19 pandemic\footnote{Universities, tech firms and 3D print enthusiasts with their own 3D printers respond to the shortage of healthcare workers' personal protective equipment (PPE) by producing it themselves.}, we demonstrate (\S\ref{sec:flexibility}) how an \name-based centralized scheduling workflow can substantially speed up PPE production by 2X without compromising existing production jobs compared to a decentralized uncoordinated approach.

We make the following contributions in the paper:
\begin{itemize}[topsep=2pt,itemsep=1pt,partopsep=2pt, parsep=2pt]
    \item We design, implement and open-source\footnote{https://github.com/anonymous/anonymous.git} \name, an \textbf{extensible, data-driven} framework for enabling centralized control in cyber-physical systems. \name achieves extensibility through plug-and-play of apps with no controller adaptations or modifications. \name achieves generality through its data-driven abstractions, allowing it to easily integrate with different CPS.\looseness=-1
    \item We design and implement a lightweight client-side library called Data Ubiquity Engine (DUE) that (1) provides \textbf{well-defined APIs} to support both \textbf{event-driven} and \textbf{data-driven} programming patterns, (2) transparently enforces on-demand data production, (3) intelligently performs batching and buffering, and (4) silently monitors application health. DUE enables easy onboarding of heterogeneous apps in \name, delivers over 100K msgs/s and incurs less than 100us delay.
    \item We demonstrate how easy it is to enable centralized control using \name (either through developing new CPS applications or integrating existing ones) in our case studies of manufacturing and networking CPS. In particular, we carry out the first study on digital-twin-equipped centralized control of additive manufacturing fleets and show that it shortens normal production time by up to nearly 40\%, and PPE production time by more than 50\% compared to a decentralized baseline approach.
\end{itemize}


\vspace{-3mm}
\section{Background}

\vspace{-2mm}

\subsection{Software-Defined Networks}
\label{sec:sdn}
The Software-Defined Networks (SDN) paradigm, in essence, describes a centralized control architecture where applications (the S in SDN) possess the intelligence of the system and fulfill many roles such as computing, decision making, and reconfiguration (of network devices) while leveraging the global view provided by a (logically) centralized controller. Compared to traditional distributed approaches, centralized solutions substantially improve the flexibility and efficiency of device management, simplify and speed up software development and iteration, and open up exciting opportunities for network management thanks to global visibility.

\vspace{-3mm}

{\flushleft\textbf{Legacy SDN controllers are not for CPS apps.}} A variety of SDN controllers are available in different programming languages, such as Ryu\cite{url:ryu}, Floodlight\cite{url:floodlight}, OpenDaylight\cite{medved2014opendaylight} and ONOS\cite{url:onos}. Legacy SDN controllers are designed as monolithic ecosystems where internal applications are mostly written in the same programming language and run alongside each other~(\ie~scaling is achieved through multiplying controller instances rather than apps). Protocols~(\eg~OpenFlow\cite{mckeown2008openflow}), device abstractions~(\eg~switches) and data types~(\eg~packets) are tailored to networking apps. Moreover, apps produce and consume data (which typically come in the form of events) in a publish-subscribe fashion.
\vspace{-3mm}

\subsection{Cyber-Physical Systems}
\label{sec:cps}
In this paper, we refer to cyber-physical systems (CPS) as integration of computation, networking, and physical processes\cite{url:cps}. Without loss of generality, CPS physical components are configurable by software, while data from the physical processes can be fed back into software for analysis and decision making. Many real-world systems are CPS, for example, smart manufacturing, internet-of-things, and autonomous vehicles.

\vspace{-2mm}

{\flushleft\textbf{Digital Twin.}} Simply put, digital twins are software replicas of physical machines that keep track of their real-time information such as status, different states, physical properties,~\etc Digital Twin is an emerging technology in CPS and a key enabler for visions like Industry 4.0\cite{qamsane2019unified,balta2019digital}. An SDN-like centralized architecture is crucial for digital-twin-based solutions because of its capability to aggregate data from different machines that digital twins heavily rely on.

\vspace{-3mm}

\subsection{Related Work}
\label{sec:related}

\textbf{SDN-like control architectures in CPS.} Previous works have explored SDN-like centralized control in many different CPS such as smart manufacturing\cite{lopez2018sdc}, IoT\cite{liu2015sdiot}, autonomous automobile\cite{jindal2018sedative,liu2016vanet}, and storage\cite{thereska2013ioflow}. These works focus on addressing specific challenges in each domain in a centralized fashion, and rely on either simulations or proprietary prototypes for proof-of-concept. \name is an open-source framework that enables researchers and developers to easily build such centralized controllers for CPS.

\textbf{Data-driven controller design.} In ADD\cite{lin2019add}, Lin~\etal present a case study of manufacturing systems and applications and identify the deficiencies of legacy SDN controller designs in handling data-driven use cases. They propose a data-driven controller design that aims to address those deficiencies, with limited implementation and evaluation. SDNator does share ADD's vision regarding the importance of supporting data-driven use cases in a controller, but SDNator has a  drastically different architecture compared to ADD and SDNator is fully implemented, well tested and open-source.

\textbf{Hierarchical and distributed controller designs.} Previous studies such as SoftMoW\cite{moradi2014softmow}, ElastiCon\cite{dixit2014elasticon}, and WE-Bridge\cite{lin2015west} improve SDN controller scalability by arranging multiple controller instances in hierarchical or distributed manners. SDCPS\cite{kathiravelu2019sd} uses message-oriented-middleware (MOM) to enable inter-domain communication between multiple CPS controller instances. Besides not addressing data-driven use cases like \name does, these works also differ from \name in their architecture: \name is not a monolithic controller or a cluster of such controllers; it is a cluster of applications loosely connected by its data backends, therefore the applications can scale independently (detailed below).




\section{Design}
\label{sec:design}

\vspace{-3mm}

In this section, we introduce the design of \name. We first establish \name's design principles, then give an overview of its architecture and detailed descriptions of its core components. We will also highlight \name's on-demand data production and fault tolerance features.

\vspace{-3mm}

\subsection{Design Overview}

\vspace{-1mm}

Compared to traditional network systems, CPS have some unique characteristics\cite{kathiravelu2019sd,lin2019add} that must be taken into account when designing \name:
\begin{enumerate}[label=(C\arabic*),topsep=2pt,itemsep=1pt,partopsep=2pt, parsep=2pt, leftmargin=*]
\item CPS have highly heterogeneous machines, protocols, applications, and data.
\item CPS are often geo-distributed~(\eg~factories, sensor networks) and cross-domain coordination is common.
\item CPS applications manifest both event-driven and data-driven patterns.
\end{enumerate}

\begin{figure}[t!]
  \centering
  \includegraphics[width=\columnwidth]{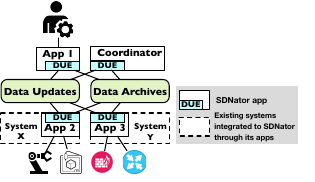}
  \vspace{-7mm}
\caption{SDNator with 3 sample applications: App~1 interfaces with users through northbound APIs; App~2 interfaces with manufacturing devices such as robot arms and 3D printers, App~3 interfaces with networking devices such as firewalls and switches.\looseness=-1}
\vspace{-3mm}
\label{fig:architecture}
\end{figure}

\vspace{-3mm}

{\flushleft An overview of \name's architecture is shown in Figure~\ref{fig:architecture}. We make the following key design decisions to accommodate the aforementioned CPS characteristics as well as to address deficiencies of legacy SDN controllers (\S\ref{sec:sdn}):}
\begin{enumerate}[label=(D\arabic*),topsep=2pt,itemsep=1pt,partopsep=2pt, parsep=2pt, leftmargin=*]
\item Applications are remotely connected to two data backends via Data Ubiquity Engine (DUE, \S\ref{sec:due}). This allows them to be language-independent~(C1), run anywhere rather than bundled with other apps, and scaled independently~(C2).
\item Apps cooperate as data consumers \& producers through a generic (C1) key-value based data schema (\S\ref{sec:data}).
\item Two data backends, Data Updates (\S\ref{sec:update}) and Data Archives (\S\ref{sec:db}), support event-driven and data-driven patterns respectively (C3).
\item A special application called Coordinator (\S\ref{sec:coordinator}) automatically handles the onboarding of new applications, and enables on-demand data production (\S\ref{sec:demand}) based on their interests and capabilities.
\end{enumerate}




\subsection{Data Backends}
\label{sec:data-store}

\subsubsection{Data Updates}
\label{sec:update}
For SDNator applications, Data Updates is the carrier of inter-application communications. It is a high-performance data distribution service that (1) allows applications to publish and subscribe to specific data items, and (2) delivers data items to subscribers in real time. Data Updates itself does \textit{not} store any information, rather, it serves the purpose of notifying interested parties of changes happening in the system (hence the name ``Updates'') and the details of those changes.


\vspace{-2mm}

\subsubsection{Data Archives}
\label{sec:db}
On the other hand, Data Archives is a mass persistent storage that stores \textit{all} information that it receives from SDNator applications. It (1) allows applications to store and retrieve any data items, and (2) supports fine-grained and range queries. Data Archives, as its name suggests, serves the purpose of persisting historical data, allowing applications to perform (big-data) analysis and make more informed decisions\cite{lin2019add}.

\vspace{-2mm}


\subsection{Applications}
\label{sec:apps}
SDNator applications, as mentioned earlier, collectively define the specific workflows and capabilities of the ``controller''. SDNator uses two simple abstractions to categorize different kinds of applications: \textbf{producer} and \textbf{consumer}, based on whether an application produces data or consumes data (or both). At initialization stage, applications need to register their \textbf{interests} (data items that they want to consume) and \textbf{capabilities} (data items that they can generate) with the Coordinator (a special application detailed in \S\ref{sec:coordinator}). After registration, an application can simply call the DUE (\S\ref{sec:due}) APIs to publish data, query historical data, or subscribe to data of interest. It is worth noting that \name does not require applications to be custom-built; rather, existing applications can be easily integrated into \name by importing the DUE library. As shown in Figure~\ref{fig:architecture}, \name applications can belong to other existing systems. Through DUE, these applications become the communication endpoints of each system. We demonstrate \name's capability to integrate existing systems in our networking case studies in \S\ref{sec:networking}.

\vspace{-2mm}


\subsection{Coordinator}
\label{sec:coordinator}

Coordinator is a special application that handles registration of new applications, monitors application heartbeats (generated by DUE, \S\ref{sec:due}) and reacts to application failures (\S\ref{sec:fault}). Coordinator sends and receives data via DUE, just like other regular applications. Its interests, quite specially, are the interests, capabilities and heartbeats of other regular applications, and its capability is assignments (specification on what data items a producer should generate and at what frequency, see \S\ref{sec:demand}). Coordinator is only involved during the onboarding process and when apps go offline/unresponsive, therefore it is not on the critical paths of any data production and consumption as they go directly to the data backends. To facilitate users and developers, Coordinator also exposes APIs for querying capabilities and statuses of existing SDNator apps.

\subsection{Data Schema and Specifications}
\label{sec:data}

In order for SDNator to accommodate applications from different systems/domains and the highly heterogeneous data they produce, there needs to be a universal and generic data representation. SDNator therefore uses key-value pairs as its data format. In our vision, the data keys must meet the following design goals:

\begin{itemize}[topsep=2pt,itemsep=1pt,partopsep=2pt, parsep=2pt]
    \item Uniquely identifies a data item
    \item Human readable
    \item Supports wildcarding
    \item Easily extendable for additional specifications
\end{itemize}

\begin{figure}[h]
  \centering
  \frame{\includegraphics[width=\columnwidth]{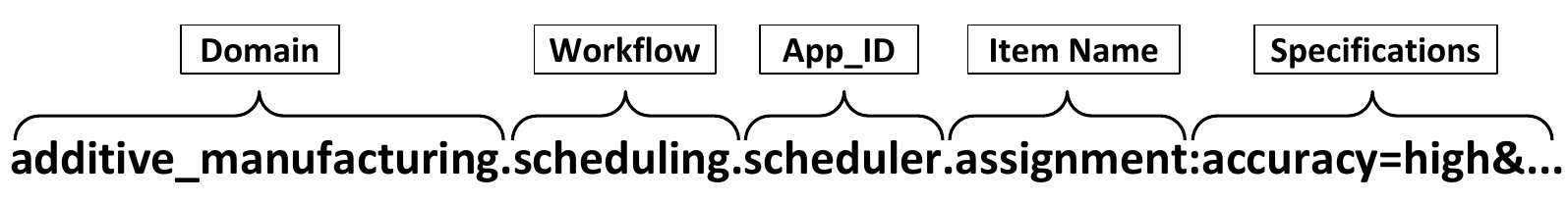}}
  \vspace{-5mm}
\caption{A sample data key of production job assignment produced by a scheduler application.}
\label{fig:key}
\end{figure}

Figure~\ref{fig:key} shows an example of SDNator's data key. Inspired by the OpenConfig\cite{url:openconfig} and gNMI\cite{url:gnmi} initiatives, which provide a vendor-neutral way of managing network devices and extracting data through a generic, hierarchical key space, SDNator adopts a hierarchical structure for its key format that meets all design goals above. Specifically, a data key is composed of several segments in increasing granularity, followed by optional specifications. The inclusion of application id allows for different applications producing the same type of data, and makes it easy to search for data produced by the same application using wildcarding.

\vspace{-3mm}
  
On top of that, we believe allowing additional specifications on data items can open up new dimensions for application-coordination and controller workflows. For instance, a machine-learning application can specify in its capability the same data item at different accuracies with a tradeoff of time.

\subsection{Data Ubiquity Engine}
\label{sec:due}
DUE is the enabler of \textbf{event-driven} and \textbf{data-driven} programming patterns in SDNator. DUE exposes APIs to SDNator applications to (1) subscribe to a data key and register a callback function with it, (2) publish new values of data keys to Data Updates, (3) persist values of data keys to Data Archives, and (4) retrieve values of data keys specified through queries.

In addition to providing data production and consumption supports, DUE also fulfills the following responsibilities:
\begin{itemize}[topsep=2pt,itemsep=1pt,partopsep=2pt, parsep=2pt]
    \item Performs handshake with the coordinator to register an application's capabilities and interests and retrieves its assignments (if a producer)
    \item Associates subscriber's callback functions with specific data keys and calls them upon Data Updates' notifications
    \item Paces write requests based on the frequency specified by the coordinator in assignments (\S\ref{sec:demand})
    \item Periodically sends heartbeat signals to the coordinator for application health monitoring (\S\ref{sec:fault})
\end{itemize}

We detail our implementation of DUE and how we address several technical challenges to improve application performance in \S\ref{sec:proto}.





\subsection{On-Demand Data Production}
\label{sec:demand}

\begin{figure}[t!]
  \centering
  \includegraphics[width=\columnwidth]{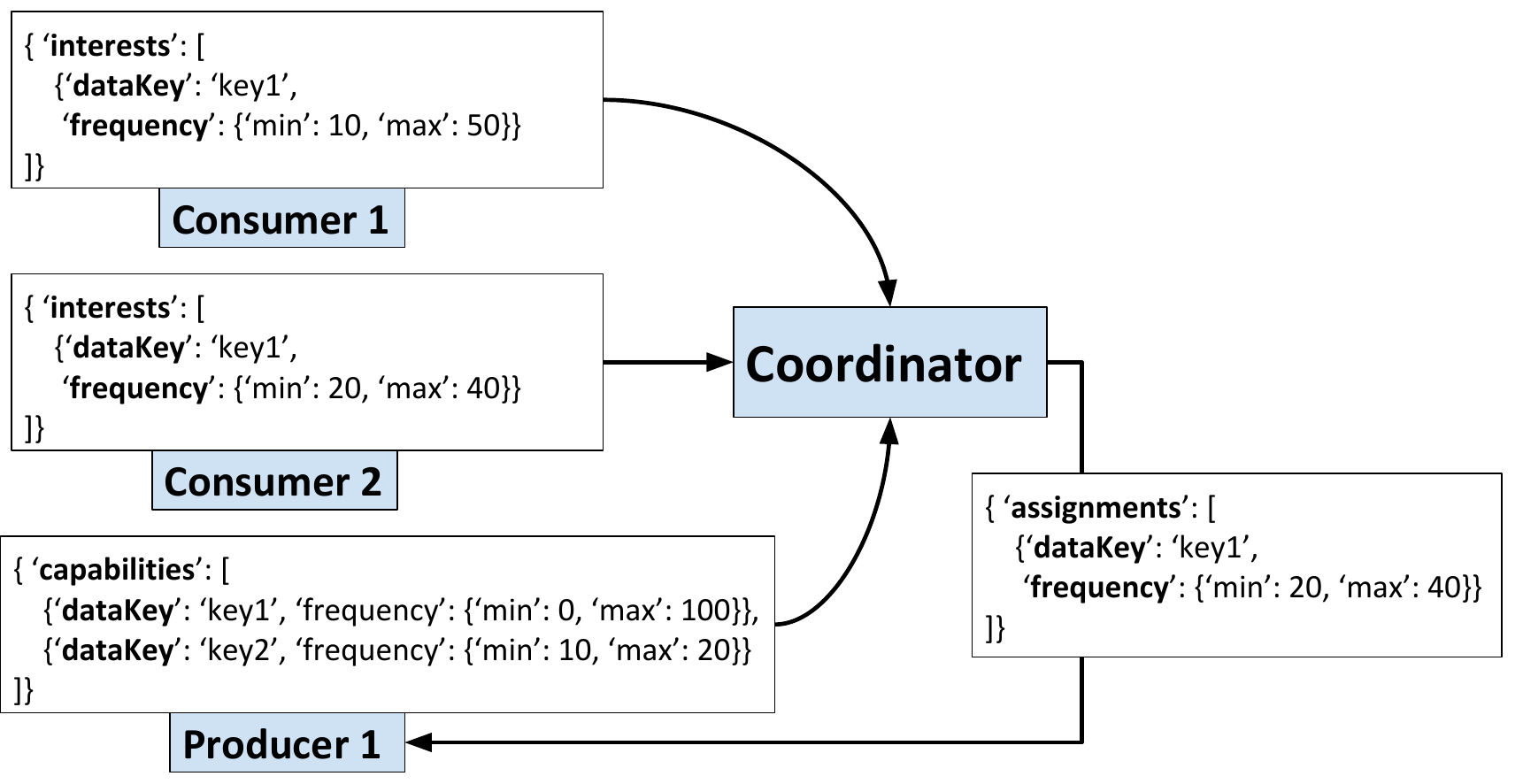}
  \vspace{-5mm}
\caption{A simple example showing the coordinator matching capabilities with interests and generating assignments with adjusted frequencies associated with each data key.}
\label{fig:consolidation}
\end{figure}

Previous study (ADD\cite{lin2019add}) showed that on-demand data production can effectively reduce bandwidth consumption on the southbound channel and latency on data requests in an SDN controller. In SDNator, considering its ``flat'' architecture, we expand this idea of on-demand data production to all applications.

\vspace{-1mm}

{\flushleft\textbf{Interest consolidation happens automatically in coordinator}}. As mentioned earlier, new applications need to register their interests and capabilities with the coordinator during onboarding. Along with the interests and capabilities, applications can also specify \textit{frequencies}, which indicate the rate at which they want to consume or can produce certain data keys. The coordinator, whenever an application joins or leaves, automatically matches all interests and capabilities on a per-data-key basis, and identifies an appropriate frequency (range) for each data key that satisfies the interested consumers within the producer's capacity. Capabilities that are matched by interests, along with (if any) the adjusted frequencies, are sent back to the corresponding producers as assignments. Figure~\ref{fig:consolidation} shows a simple example of the above process.

\vspace{-1mm}

{\flushleft\textbf{Frequency enforcement happens transparently and automatically in DUE}}. Upon receiving the assignments, DUE automatically parses the them and records the frequency (range) associated with each data key. Whenever a producer calls DUE API to publish a data key, DUE checks whether that data-key is frequency-bound, and paces that write if necessary to meet the frequency specification.

Through on-demand production, SDNator eliminates redundancy in data production (by consolidating all interests) while offering more flexibility and control (See \S\ref{sec:sonata}).

\vspace{-2mm}

\subsection{Fault Tolerance, Detection and Recovery}
\label{sec:fault}
We only consider two sources of faults/failures in SDNator\footnote{Detecting faults or incorrectness in data is beyond the scope of \name's capabilities and should/can be handled by consumer applications.}: the data backends and applications. For data backends‘ reliability, SDNator relies on modern data stores'\cite{url:redis-cluster,url:mongo-replication,url:kafka} fault tolerance mechanisms without assuming any better or worse. Here we focus on how SDNator handles failures of applications.

\vspace{-2mm}

{\flushleft\textbf{Fault detection is achieved through periodic heartbeat from DUE to coordinator.}}
As mentioned earlier, DUE periodically sends heartbeat signals to the coordinator, which allows the coordinator to keep track of the statuses of each application: be it offline, unresponsive, or simply just idle. This process is accomplished, similar to interest and capability registration, by coordinator subscribing to all heartbeat channels of each application. This is feasible based on two key points: (1) Data Updates is publish-subscribe based, therefore sending heartbeat is non-blocking; (2) aggregate heartbeat rate, even with 100 applications each sending at 10 times per second, will still be orders of magnitude lower than full capacity of an SDNator consumer (see \S\ref{sec:benchmark}).

\vspace{-2mm}

{\flushleft\textbf{Fault tolerance is achieved through redundancy in application instances.}}
In SDNator, the same application can be launched multiple times for redundancy. DUE internally maintains an instance-specific UUID to differentiate between instances. Different instances will register and onboard like any new applications, except that coordinator will associate them with the same application id. Coordinator will randomly\footnote{Different selection strategies can be easily implemented. For instance, pick instances that have the lowest latency or most resources.} select an active instance when generating the first assignments, and sticks to that instance until it goes offline or unresponsive.

\vspace{-2mm}

{\flushleft\textbf{Fault tolerance and detection allow for fast fault recovery.}}
With fault detection, coordinator can quickly perceive when applications go offline or unresponsive\footnote{DUE sends a special heartbeat when application goes offline, whereas heartbeat timeout indicates unresponsiveness.}. And because redundancy exists, coordinator can easily switch to another active instance in no time to quickly recover from failures. It is worth noting that coordinator itself is not on the critical path of data production or consumption; even if coordinator crashes, it can simply restart and recover its states from new heartbeat information or previous checkpoints in Data Archives, as it only contains \textit{soft state} information collected from DUE applications.

\vspace{-3mm}

\section{Implementation and Technical Challenges}
\label{sec:proto}
In this section, we describe our prototype of SDNator including how we implement Data Ubiquity Engine~(\S\ref{sec:proto-due}) and the frameworks used for Data Updates~(\S\ref{sec:redis}) and Data Archives~(\S\ref{sec:mongo}). We also introduce several optimization techniques we adopt in DUE to address technical challenges that emerge during implementation~(\S\ref{sec:optmization}).

\vspace{-3mm}

\subsection{Data Ubiquity Engine}
\label{sec:proto-due}
We implement Data Ubiquity Engine (DUE) as a software library in Python\footnote{for rapid prototyping and decent performance as shown in \S\ref{sec:benchmark}} supporting versions 2.7+ and 3.7+, using fewer than 2000 lines of code. DUE can be easily imported in any user applications written in the above environments. The dependencies and development patterns used are readily available and reproducible in other common languages like Java and C++.

The DUE library is composed of three major components: 1) the DUE API that provides applications with generic event-driven and data-driven programming patterns, 2) a Pub-Sub driver that implements the event-driven APIs using Redis Publish-Subscribe functions\cite{url:redis-pubsub}, and 3) a Database driver that implements the data-driven APIs using MongoDB\cite{url:mongo}.


Being a lightweight client-side library, DUE does not require an adaptation in programming styles like MapReduce as in Hadoop\cite{url:hadoop} and RDD as in Spark\cite{url:spark}, nor does it require complex interactions with job scheduling backends such as YARN\cite{url:yarn} (although it can be integrated into SDNator for resource scheduling). Therefore, it is much easier for developers to on-board their applications, for example, by simply including the following few lines of code in their existing applications:

\begin{lstlisting}[belowskip=-2\baselineskip,language=Python]
from sdnator_due import *
# Optional, configure the Data Archives backend
due.set_db(...)
# Optional, configure the Data Updates backend
due.set_pubsub(...)
due.init("MyNewAppId", PRODUCER | CONSUMER)
\end{lstlisting}

\subsection{Data Updates}
\label{sec:redis}
For implementing Data Updates, we choose Redis\cite{url:redis} for its high performance and ease of use, as it was used as the message broker backend for some very high throughput and low latency messaging services like Pusher\cite{url:pusher}. Specifically, we use Redis' Publish-Subscribe\cite{url:redis-pubsub} functionality to implement  real-time and in-order message delivery, and the event-driven programming interface.

For an SDNator application that wants to publish a new value for a data-key, it can simply:
\begin{lstlisting}[belowskip=-1.5\baselineskip,language=Python]
# specify PUB_ONLY to bypass Data Archives
due.write(dataKay, dataValue, [PUB_ONLY])
\end{lstlisting}

For an SDNator application that wants to subscribe to a data-key and register a callback function, it can simply:
\begin{lstlisting}[belowskip=-1.5\baselineskip,language=Python]
observer = due.observe(dataKey)
observer.subscribe(lambda data: call_back_func(data))
\end{lstlisting}

The \verb|observer| object returned from \verb|due.observe| is an RxPy\cite{url:rxpy} Subject with powerful data stream capabilities inherited from the ReactiveX programming paradigm\cite{url:reactivex}.

Redis uses TCP for its connections, which has huge implications for SDNator applications: even a modest level of network latency can severely reduce application throughput by limiting the number of communication round trips per second. We address this challenge in \S\ref{sec:optmization}.

One potential drawback of using Redis is the lack of delivery guarantee, which is a trade off for its high performance and simplicity. Although we don't observe any losses in our benchmarks~(\S\ref{sec:benchmark}), if needed, the Data Updates backend can be easily swapped with a reliable message broker like Kafka\cite{url:kafka}, thanks to DUE's encapsulation.

\subsection{Data Archives}
\label{sec:mongo}
For implementing Data Archives, we choose MongoDB\cite{url:mongo} for its relatively good performance and rich feature set for data queries.

When producers call \verb|due.write()|, DUE automatically executes writes to both Data Updates and Data Archives unless otherwise specified~(\eg the \verb|PUB_ONLY| flag shown above). Besides the data key and value, we also include timestamp and application id for the writes to MongoDB, which would facilitate applications such as the coordinator to query data by producer or by time range. Since Data Archives can be slower in write speed, this may cause a slowdown for Data Updates and therefore application throughput. We address this challenge in \S\ref{sec:optmization}.

For an SDNator application that wants to retrieve historical data from Data Archives:
\begin{lstlisting}[belowskip=-1.5\baselineskip,language=Python]
# fetch all historical records of 'dataKey'
data = due.get(dataKey)
# or use advanced MongoDB queries
data = due.get({
    'dataKey' : dataKey,
    'timestamp': {
        '$gt': someDate
    }
})
\end{lstlisting}

Similarly, because DUE's APIs are backend-independent, MongoDB can be swapped with other mass storage databases.

\subsection{Technical Challenges}
\label{sec:optmization}

\subsubsection{Network Latencies}
As mentioned earlier, network latencies can severely reduce application throughput. As described in \S\ref{sec:update}, message ordering needs to be strictly preserved for Data Updates, therefore multi-threading is not an option. Since network latency impacts application throughput, by limiting the number of (TCP) round trips, we can improve throughput by increasing the data volume in each trip (\ie multiplexing). Since Redis natively supports multiplexing and demultiplexing through its pipelining\cite{url:redis-pipeline} feature, DUE only needs to process messages in batches and provide internal buffering. To make it more instrumental and flexible, we allow developers to configure their own batch sizes depending on network conditions and communication patterns. Batching can improve application throughput by up to 5X with little latency penalties, as demonstrated in \S\ref{sec:throughput}.

\subsubsection{Slower Data Archives Writes}
Data Archives as a mass persistent storage cannot deliver the same write speeds as Data Updates. In order to address this performance gap, we adopt the following two strategies:

\textbf{Make Data Archives writes non-blocking.} MongoDB client writes in a blocking fashion. To go around that, we first resolve to Thread Pool. We notice during testing, however, that Python's Global Interpreter Lock (GIL)\cite{url:gil} causes severe performance degradation. Therefore, we switch to Process Pool instead. Though multiprocessing inherently carries the overhead of IPC and data copying which is not ideal, we see in benchmarks~(\S\ref{sec:throughput}) it delivers satisfying performance. Since this is caused by an inherent limitation of the Python interpreter, it's also safe to assume that our benchmark results serve as a \textbf{lowerbound} of DUE's performance.

\textbf{Make Data Archives writes more efficient.} Similar to how we optimize Data Updates writes to account for network latencies, we adopt batching and buffering for Data Archives writes as well. \S\ref{sec:throughput} shows results of the improvement achieved through batching.

It is also worth noting that, with Process or Thread Pools, the order of the Data Archives writes won't be guaranteed. To mitigate this, DUE silently appends a timestamp to each data item being written to Data Archives so that order can be restored during future queries. And to make queries faster, we have adopted common database optimization techniques like indexing\cite{url:index}. Applications are allowed to attach a \verb|index: true| to their capabilities~(\S\ref{sec:coordinator}), so any \verb|due.get()| requests toward those items can be performed much faster.

\vspace{-2mm}
\section{Benchmarks}
\label{sec:benchmark}

\vspace{-2mm}

In this section, we demonstrate SDNator's performance (Figures~\ref{fig:due-delay}, \ref{fig:vs-redis}, \ref{fig:mongo-opt}) and scalability (Figures~\ref{fig:rtt-ave}, \ref{fig:rtt-sum}) through a series of benchmarks. These benchmark results help quantitatively evaluate the latency overhead, achievable application throughput, and scalability of the system. Unless otherwise specified, benchmarks are run on an Ubuntu 18.04 (Linux 4.15.0) machine with Intel Core i7-7700K@4.2GHz quad-core processor and 32GB of RAM. Each experiment is repeated ten times.

\subsection{End-to-end Latency}
\label{sec:latency}

\begin{figure}[t!]
    \centering
    \includegraphics[width=0.9\columnwidth]{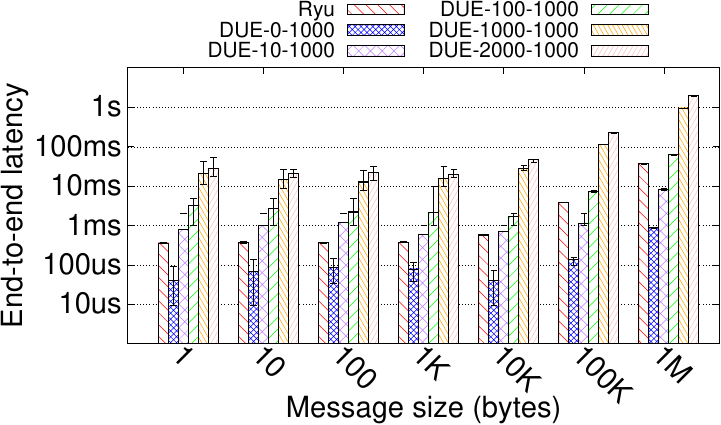}
    \vspace{-2mm}
    \caption{A comparison between Ryu and SDNator on end-to-end latency between a producer and a consumer when sending messages at different sizes. For SDNator: \\DUE-\{Data Updates batch\}-\{Data Archives batch\}.}
    \label{fig:due-delay}
\end{figure}

Unlike existing SDN controllers\cite{url:ryu,medved2014opendaylight,url:onos,url:floodlight} where applications (especially built-in services like \textit{topology discovery} and \textit{routing}) typically communicate via inter-procedure calls, SDNator uses remote-procedure calls and introduces a path stretch through the Data Updates~(\S\ref{sec:update}). In order to understand the latency ``penalty'' potentially imposed by SDNator's design, we measure the inherent\footnote{when communicating through the loopback interface of the host machine} end-to-end latency~(\ie duration between a batch of messages is sent and received) between two SDNator applications when sending messages of different sizes\footnote{For all experiments, we randomize the bytes in each message to eliminate potential caching influences.} using different batching strategies. For comparison, we also include results from Ryu\cite{url:ryu}, a widely used SDN controller written in Python. We write two applications in Ryu and leverage Ryu's own event APIs to publish and subscribe to our custom events that contain those messages of various sizes. The results are shown in Figure~\ref{fig:due-delay}.\looseness=-1

From the figure we can see that SDNator's base latency (when no batching is enabled for Data Updates) is even lower than Ryu's. Even at a batch size of 100, SDNator is still comparable to Ryu.

\begin{tcolorbox}[boxsep=-2pt,after skip=4pt]
\textbf{Finding 1.} SDNator has a lower inherent latency footprint than Ryu. Even with moderate batching enabled, SDNator can still deliver comparable performance.
\end{tcolorbox}

It is worth noting that we have introduced a \verb|NO_WAIT| keyword to \verb|due.write()| and a flushing API such that writes can be executed right away even when batching is enabled to further reduce unnecessary latency overheads.

\subsection{Application Throughput}
\label{sec:throughput}
Achieving high throughput between applications is both critical and challenging for SDNator. As described in \S\ref{sec:optmization}, Data Ubiquity Engine (DUE) adopts techniques like batching and buffering to improve application throughput. How effective are these techniques? How much additional overhead does DUE incur? To answer these questions, we conduct several experiments (same setup as above, measuring throughput instead of latency) with our prototype using different DUE configurations detailed below:

\vspace{-2mm}

\subsubsection{Data Updates Batching}
\begin{figure}[t!]
    \centering
    \includegraphics[width=0.9\columnwidth]{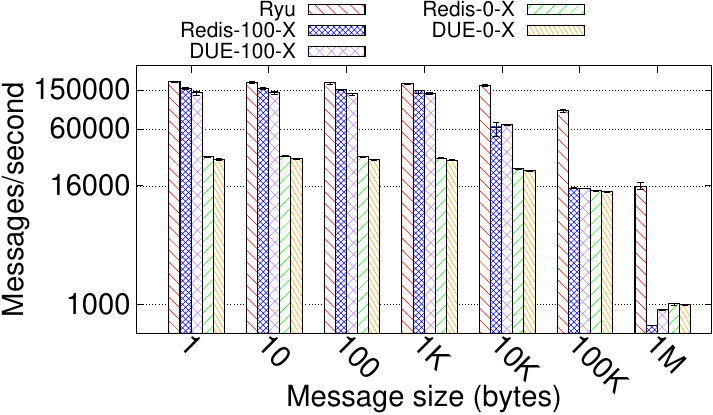}
    \vspace{-2mm}
    \caption{Application throughput when using Ryu, raw Redis APIs w/ and w/o batching, and DUE w/ and w/o batching; Data Archives is disabled.}
    \label{fig:vs-redis}
\end{figure}

In the first experiment, we focus on the effectiveness of Data Updates batching and the overhead of DUE. We measure and compare application throughput (in messages per second) when using Ryu, raw Redis pub-sub APIs with and without batching, and DUE with and without batching. Based on results in Figure~\ref{fig:due-delay}, batch size is set to 100 for both Redis and DUE. Results are shown in Figure~\ref{fig:vs-redis}.

We can clearly see that DUE+Redis performs fairly close to raw Redis at batch size 100. In fact, the difference is less than 10\%. We can also tell that batching has up to 5X improvement on throughput. Ryu does deliver a higher throughput than Redis and Redis+DUE, but the difference is marginal until the message size climbs up to more than 10KB when Redis is bound by bandwidth of the loopback interface while Ryu is mostly bound by memory speed. In Ryu, events typically carry packets that are less than 1.5KB, therefore SDNator is fairly close.

\begin{tcolorbox}[boxsep=-2pt,after skip=0pt]
\textbf{Finding~2.} DUE incurs 10\% or less overhead compared to raw Redis APIs.\\
\textbf{Finding~3.} A Data Updates batch size of 100 can improve app throughput by up to 5X in both DUE \& Redis.\\
\textbf{Finding~4.} Ryu delivers higher application throughput than SDNator, although the gap only becomes substantial when message sizes are 10KB+.
\end{tcolorbox}

\subsubsection{Data Archives Batching}
\begin{figure}[t]
    \centering
    \includegraphics[width=0.9\columnwidth]{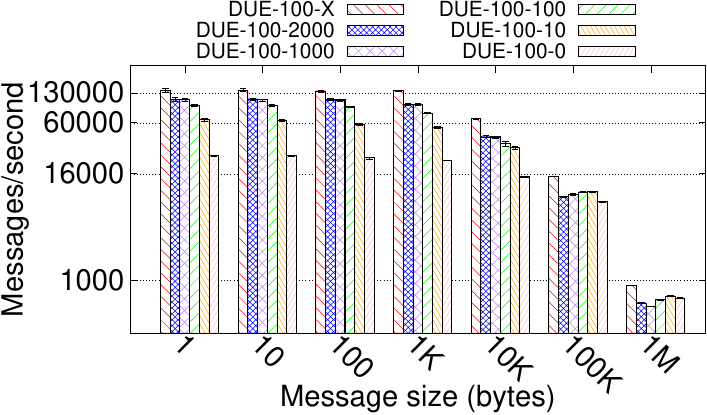}
    \vspace{-2mm}
    \caption{Application throughput w/ different Data Archives batch sizes; Data Updates batch size set to 100.}
    \label{fig:mongo-opt}
\end{figure}

In the second experiment, we focus on the overhead of Data Archives (given it is slower than Data Updates) and the effectiveness of Data Archives batching. We measure and compare the application throughput with different Data Archives batch sizes, as shown in Figure~\ref{fig:mongo-opt}.

\begin{tcolorbox}[boxsep=-2pt,after skip=2pt]
\textbf{Finding~5.} Enabling Data Archives slows down application throughput by \textasciitilde20\% to \textasciitilde80\% depending on the Data Archives batch size.\\
\textbf{Finding~6.} A Data Archives batch size of 1000 can improve application throughput by up to 3.4X.
\end{tcolorbox}

\subsection{Scalability}
\label{sec:scalability}

\begin{figure}[t!]
    \centering
    \includegraphics[width=0.8\columnwidth]{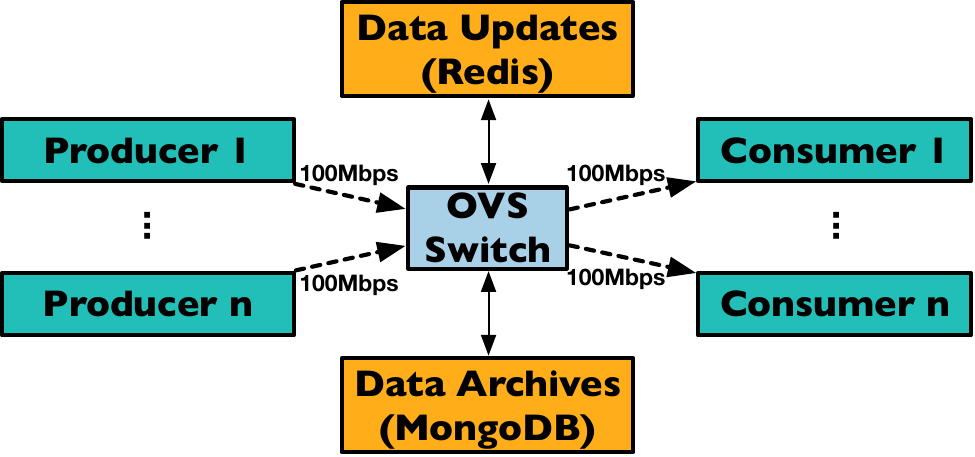}
    \vspace{-3mm}
    \caption{Mininet emulated topology described in \S\ref{sec:scalability}. Links marked with dotted arrows have 100Mbps bandwidth and latency from 0 to 50ms.}
    \label{fig:topology}
\end{figure}

\begin{figure}[t!]
    \centering
    \includegraphics[width=0.8\columnwidth]{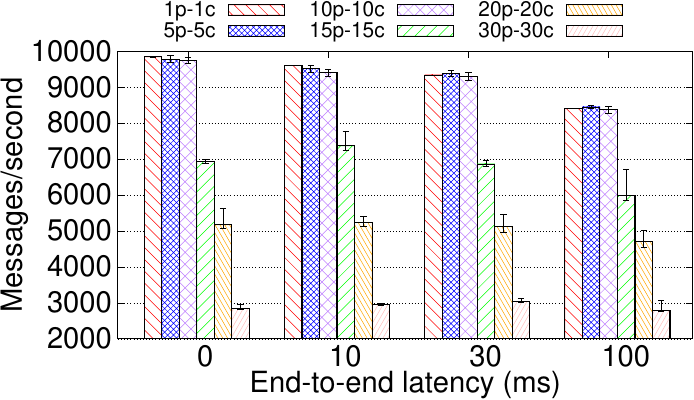}
    \vspace{-3mm}
    \caption{Individual application throughput under different network latencies and \# of consumer/producer pairs; Data Archives batch size set to 12500.}
    \label{fig:rtt-ave}
\end{figure}

\begin{figure}[t!]
    \centering
    \includegraphics[width=0.8\columnwidth]{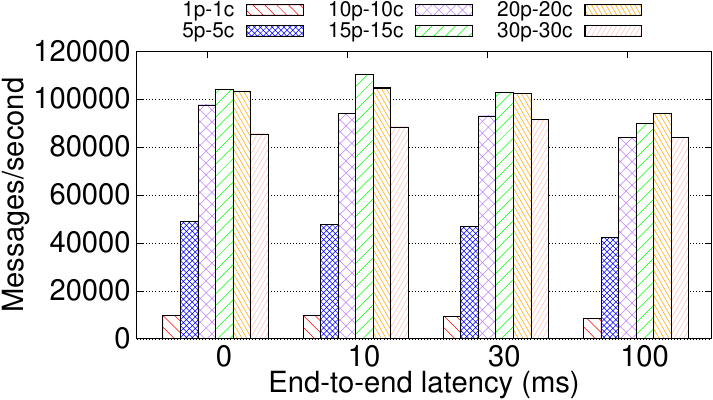}
    \vspace{-3mm}
    \caption{Aggregate application throughput under different network latencies and \# of consumer/producer pairs; Data Archives batch size set to 12500.}
    \label{fig:rtt-sum}
\end{figure}

To further test SDNator in a more realistic setting with network latencies and evaluate its scalability, we carry out another experiment\footnote{This experiment is run on an Ubuntu 18.04 (Linux 4.15.0) server with Intel Xeon E5-4620v2@2.60GHz 8-core processor and 128GB of RAM.} in which we use Mininet\cite{url:mininet} to emulate a star topology with applications running on different hosts with 100Mbps bandwidth, as shown in Figure~\ref{fig:topology}. The message size is fixed at 1000 bytes and Data Updates batch size 12500. The number of pairs of producers and consumers is from 1 to 30, and the one-way latency between producer/consumers and the switch varies from 0 to 50ms (\ie end-to-end latency from 0 to 100ms). We measure individual application throughput on each host to see if they are fairly close, and aggregate them to see if overall performance suffers as contention increases. Results are shown in Figures~\ref{fig:rtt-ave} and \ref{fig:rtt-sum} for individual and aggregate throughput respectively.

If we look at the results for up to 15 producer/consumers in Figure~\ref{fig:rtt-ave}, when individual throughput is only capped by bandwidth, we can clearly see that (1) higher lantencies incur moderate drop on throughput, and (2) more applications does not lead to either lower throughput or high fluctuations in throughput. As we move to Figure~\ref{fig:rtt-sum}, we can further tell that aggregate throughput does not drop even with 40 applications. Even with 60 applications, we only see a slight drop in aggregate throughput likely due to limited computing resources of our machine.

Thanks to SDNator's modular design and the incorporation of state-of-the-art data stores, its scalability can be further improved through replication\cite{url:mongo-replication} and clustering\cite{url:redis-cluster} which are common practices today. For example, operators managing geo-distributed sites can opt for running local instances of data backends to reduce the impact of network latencies and let the backends' clustering and replication mechanisms take care of the synchronization.

\begin{tcolorbox}[boxsep=-2pt,after skip=0pt]
\textbf{Finding~7.} Higher network latencies lead to slightly lower application throughput; however, even with 100ms end-to-end latency, individual applications can still achieve \textasciitilde67\% utilization of 100Mbps bandwidth.\\
\textbf{Finding~8.} Contention does not cause inequalities or degradation among SDNator applications\\
\textbf{Finding~9.} SDNator scales well to a fairly large number of applications without noticeable degradation.
\end{tcolorbox}


\section{Case Studies}
\label{sec:cases}


\subsection{Additive Manufacturing}
\label{sec:manufacturing}



As mentioned earlier (\S\ref{sec:related}), the idea of centralized control has been explored in the context of smart manufacturing which relies heavily on cyber-physical systems (CPS) for their reconfigurability and data availability. Currently, no existing centralized control framework for CPS is available to allow researchers or developers to easily test out their applications. In this case study, we focus on one category of smart manufacturing: additive manufacturing, \aka \textbf{3D printing}. More specifically, we perform the first-ever study of digital-twin-based scheduling\footnote{The scheduling logic for the production has a great impact on the makespans (\ie time from production order to finished product).} for an additive manufacturing fleet\footnote{A printer fleet consists of a number of 3D printers that print 3D objects according to assigned jobs}\cite{balta2018centralized} by leveraging SDNator to build centralized control workflows for multiple 3D printing systems. Our study clearly suggests that a digital-twin-equipped centralized controller helps reduce production time (\S\ref{sec:productivity}) as well as detect and react to anomalies in real time (\S\ref{sec:reliability}). Moreover, we evaluate our \name-based controller in a more complex scenario where urgent requests such PPE\cite{url:citizen-ppe} or ventilators\cite{url:ventilator} are added on top of existing production jobs and demonstrate that it adjusts production plans on the fly to speed up PPE production by 2X without compromising existing production jobs (\S\ref{sec:flexibility}).

\subsubsection{An SDNator-based Controller for 3D Printing}



\begin{figure}[t!]
    \centering
    \includegraphics[width=0.7\columnwidth]{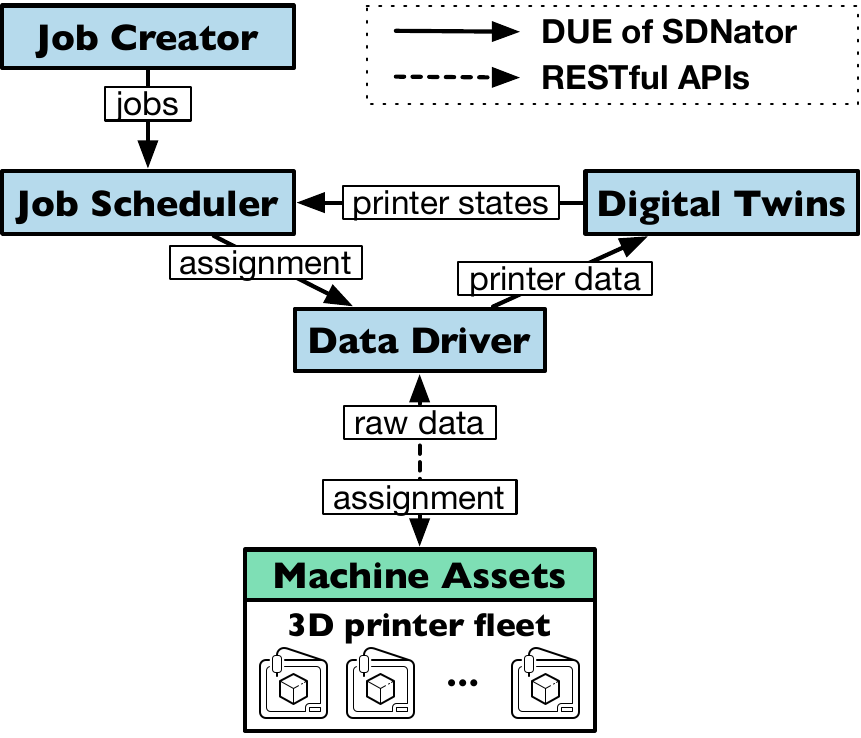}
    \vspace{-2mm}
    \caption{An SDNator-based centralized control workflow for additive manufacturing scheduling. Each blue box is an \name app. The Job Scheduler subscribes to different amount of information from Digital Twins to implement different scheduling algorithms.}
    \label{fig:scheduling}
\end{figure}
Figure~\ref{fig:scheduling} shows the workflow of the central controller built using SDNator, which consists primarily of the following \name apps:

\vspace{-2mm}

{\flushleft\textbf{Job Creator}} initializes the production orders (both regular jobs and PPE jobs) at predefined times.


{\flushleft\textbf{Job Scheduler}} receives the production orders from the Job Creator and the state information of the fleet from the Digital Twins (DT) app and executes a scheduling algorithm (described below) to assign jobs to machines in the fleet.

\vspace{-2mm}

{\flushleft\textbf{Digital Twins.}} A machine-state DT monitors the functional state (\eg idle, running, \etc), physical properties, and time information of a printer. A queue DT monitors the job queue and updates the queue finish time, queue length, average queue wait, and queue history. An anomaly detection DT monitors the sensors on the printer to detect anomalies, and reports any anomalies to a fleet DT. When there is an anomaly, the fleet DT schedules a maintenance event, which stops the current job to save time and material and the current job is rescheduled according to the scheduling algorithm. The DT app monitors each printer, sensor, and queue to update machine-states, queue, and anomaly detection DTs.

\vspace{-2mm}

{\flushleft\textbf{Data Driver}} sends the job assignments to the AM fleet through a \textbf{Machine Assets} app that faithfully emulates the Jedi RESTful API of Ultimaker 3\cite{url:ultimaker} Fused Deposition Modeling (FDM) printers, which are off-the-shelf printers commonly used in practice. The Data Driver also transfers data from the Machine Assets to the DT app so that DTs maintain the up-to-date state of the printers. The Machine Assets manages an emulated fleet of 3D printers 
modeled\cite{balta2019digital} based on the Ultimaker 3 FDM printer. Each printer has a job queue and two sensors for anomaly detection.




{\flushleft To demonstrate the performance benefits of centralized control for 3D Printing, we implement 4 scheduling algorithms that leverage the central controller to different extents:}

\vspace{-2mm}

{\flushleft\textbf{Decentralized}} does not utilize any global view provided by \name. Each machine takes a production order when queue is empty with no knowledge of each other. 
This is also the most common PPE production scheme during COVID-19 where contributors with existing AM capabilities produce PPEs for healthcare workers in an uncoordinated fashion.

\vspace{-2mm}

{\flushleft\textbf{Centralized-Baseline.}} The scheduler is aware of all the machines in the AM fleet thanks to \name making the fleet size captured by Data Driver available to other apps. It is however not using the states of the machines that Digital Twins are publishing. As a result, the Job Scheduler 
simply distributes the production orders to all the machines in the fleet uniformly.

\vspace{-2mm}

{\flushleft\textbf{Centralized-FCFS.}}
The Digital Twins publish realtime state information (queue, machine-state, and anomaly of each machine) for other applications to consume. 
The scheduler uses the availability and queue length of the printers to schedule each job in a production order to a machine in the fleet, in a first come, first served (FCFS) fashion.

\vspace{-2mm}

{\flushleft\textbf{Centralized-Dynamic}} minimizes the expected makespan of the jobs through optimization by making use of all the machine-state information published by the Digital Twins. Specifically, we form an integer linear program (ILP) to evaluate the optimal number of jobs to be sent to each machine based on their availability, setup time, and expected queue finish time. 
\vspace{\baselineskip}

\subsubsection{Shortening the Production Time}
\label{sec:productivity}

\begin{figure}[t!]
    \centering
    \includegraphics[width=0.7\columnwidth]{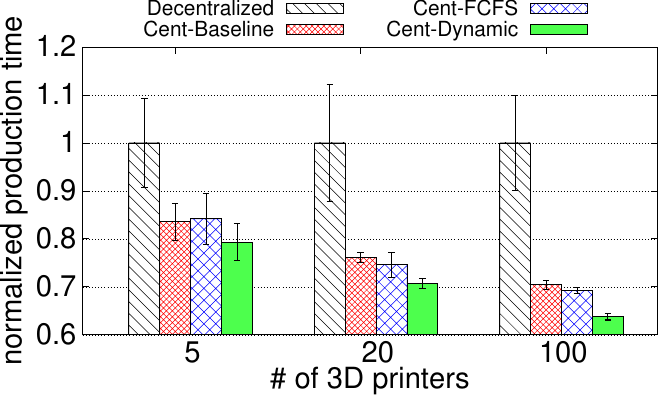}
    \vspace{-2mm}
    \caption{Comparison of normal job makespans between decentralized and centralized scheduling algorithms.}
    \label{fig:adm-reg}
\end{figure}

\begin{figure}[t!]
    \centering
    \includegraphics[width=0.7\columnwidth]{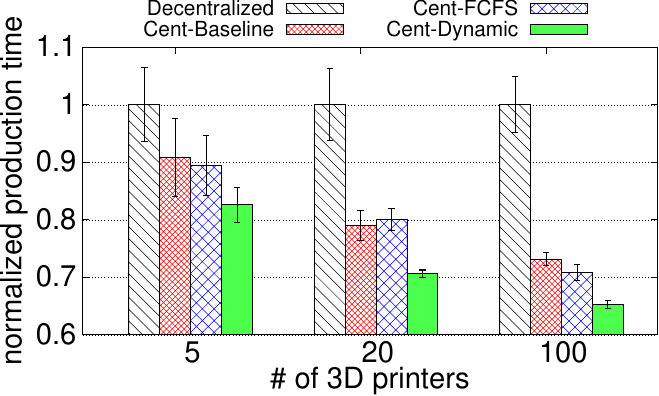}
    \vspace{-2mm}
    \caption{Comparison of job makespans between decentralized and centralized scheduling algorithms with anomalies.}
    \label{fig:adm-ad}
\end{figure}

To begin, we consider a baseline production scenario where the AM fleet executes normal orders without anomalies or demand changes. Figure~\ref{fig:adm-reg} shows the performance of each scheduling algorithm, in terms of makespans. Final results are normalized by the maximum average makespan of the worst-performing algorithm for ease of comparison. The decentralized algorithm performs the worst as expected since jobs are assigned to machines randomly. The centralized baseline solution outperforms the decentralized one and performs comparably to the centralized FCFS solution. The centralized dynamic solution excels as it optimizes job assignments to minimize expected makespan, potentially saving hours, days or even weeks.

\begin{tcolorbox}[boxsep=-2pt,after skip=0pt]
\textbf{Finding~10.} The benefit of a global-view provided by the SDNator-based central controller is evident from the baseline results. The schedulers that utilize a centralized approach can greatly improve the makespans in the baseline scenario.
\end{tcolorbox}


\vspace{\baselineskip}

\subsubsection{React To Real-Time Anomalies}
\label{sec:reliability}

Here we consider a production scenario with probabilistic random anomalies on printers while producing the same production orders as in the baseline scenario (\S\ref{sec:productivity}). If an anomaly is detected during a print, the same job needs to be re-printed.
Figure~\ref{fig:adm-ad} shows the performance of each strategy, in terms of makespans. Similar to \S\ref{sec:productivity}, the benefit of a global view is pronounced by the shorter makespans of the centralized solutions. When an anomaly is detected by the anomaly detection DT, the current job is reprinted on the same printer for the decentralized scheduler, and scheduled to (potentially) another printer based on the strategy of the centralized schedulers.

As the number of machines in the fleet increases, the benefit of the global view through the DTs becomes more evident. The run-time information provided by the DTs improves the efficiency of the dynamic scheduler at all scales.

\begin{tcolorbox}[boxsep=-2pt,after skip=0pt]
\textbf{Finding~11.} SDNator-based central controller improves reliability by enabling digital-twins that capture real-time information such as machine status and occupancy for quick reaction to anomalies and effective mitigation.
\end{tcolorbox}

\begin{figure}[t!]
    \centering
    \includegraphics[width=0.7\columnwidth]{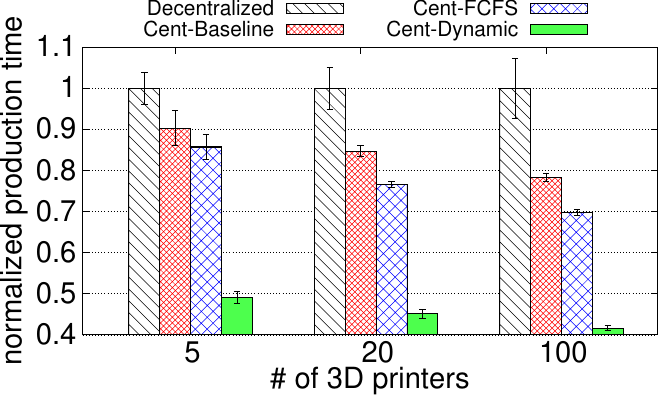}
    \vspace{-2mm}
    \caption{Comparison of PPE makespans (w/ normal jobs) between decentralized and centralized scheduling algorithms.}
    \label{fig:adm-ppe-ppe}
\end{figure}

\begin{figure}[t!]
    \centering
    \includegraphics[width=0.7\columnwidth]{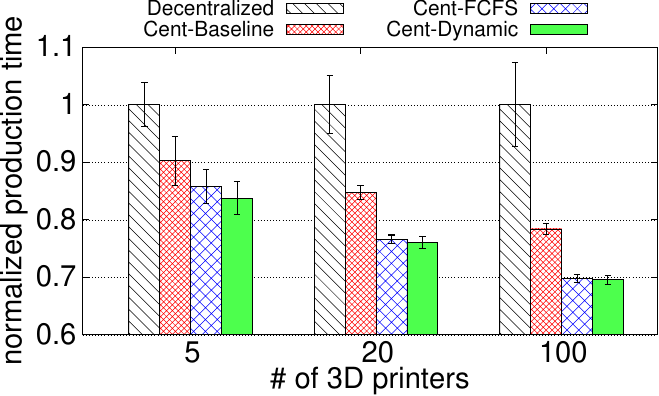}
    \vspace{-2mm}
    \caption{Comparison of final job makespans (w/ PPE) between decentralized and centralized scheduling algorithms.}
    \label{fig:adm-ppe-reg}
\end{figure}



\vspace{\baselineskip}

\subsubsection{Adjust Production Plan On The Fly}
\label{sec:flexibility}
Finally we consider a production scenario where an urgent request of Personal Protective Equipment (PPE) is received while the fleet is maintaining baseline production described in \S\ref{sec:productivity}.
Figure~\ref{fig:adm-ppe-reg} shows 
the final makespan of each scheduler 
and Fig.~\ref{fig:adm-ppe-ppe} shows the makespan of the PPE orders.

The PPE production scenario is of high importance and requires timeliness due to its wide applicability during the COVID-19 pandemic. As PPE production has high priority, it is expected that a dynamic scheduler should be able to prioritize these orders over the preexisting orders in the AM fleet. While all the other schedulers utilize a basic FIFO queue, the dynamic scheduler utilizes a reconfigurable queue where existing jobs in the queue can be preceded by priority jobs. The dynamic scheduler uses the same ILP-based makespan minimization optimization to schedule the PPEs with priority while balancing the total production makespan. The results in Figure~\ref{fig:adm-ppe-ppe} show that this approach enables the dynamic scheduler to produce the PPEs \textasciitilde50\%-100\% faster than the other schedulers, without sacrificing overall makespans as shown in Figure~\ref{fig:adm-ppe-reg}.

\begin{tcolorbox}[boxsep=-2pt,after skip=0pt]
\textbf{Finding~12.} SDNator-based central controller provides more agility to adapt to changes in production demands while preserving efficiency.
\end{tcolorbox}

\subsection{Networking}
\label{sec:networking}

To further demonstrate SDNator's generality and extensibility, we implement two control workflows in networking systems by converting existing networking applications into data producer/consumers and plug-and-play, as shown in Figure~\ref{fig:workflows}. 

\subsubsection{Network Telemetry and Monitoring}
\label{sec:sonata}
Gupta \etal\cite{gupta2018sonata} show via their Sonata framework the power and flexibility of using SDN techniques like P4\cite{bosshart2014p4} and big data systems like Spark\cite{url:spark} to perform network telemetry at scale. It occurs to us that Sonata is an ideal upstream SDNator producer to efficiently generate and stream digested information to downstream consumers. More specifically, we can interface Sonata's custom querying capability with SDNator's on-demand data production mechanism (\S\ref{sec:demand}) to allow other applications to determine what information to extract from network traffic at a fine granularity by specifying interests that can be mapped into specific queries in Sonata, \eg \verb|network.traffic_capture.sonata.tcp:flag=syn| using SDNator's data schema (\S\ref{sec:data}).

\begin{figure}[t!]
    \centering
    \includegraphics[width=0.7\columnwidth]{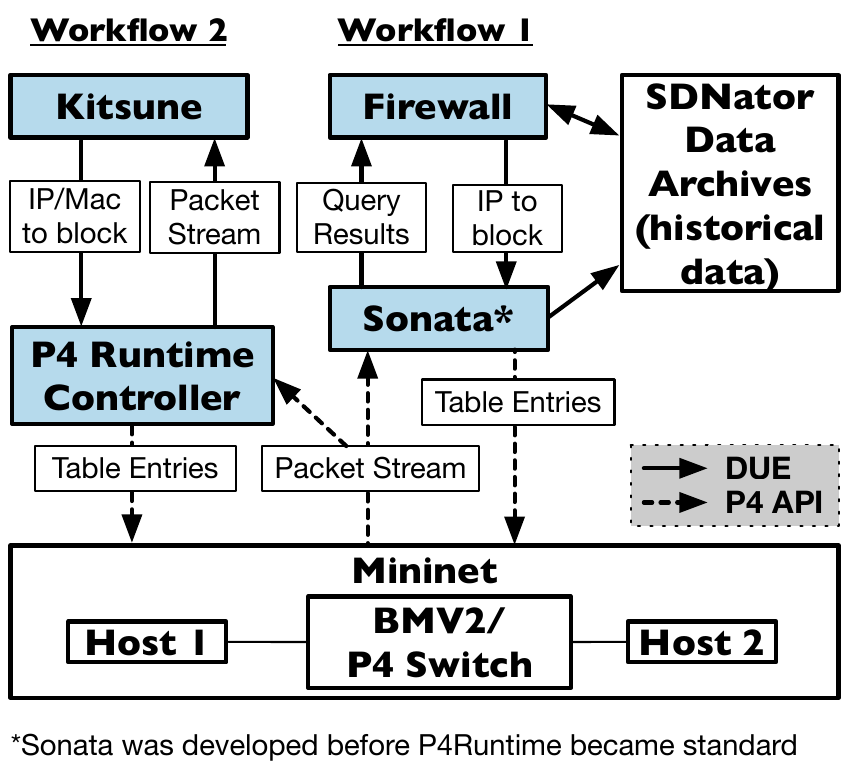}
    \vspace{-2mm}
    \caption{Control workflows constructed by using SDNator to integrate different existing networking systems.}
    \label{fig:workflows}
\end{figure}

By simply importing DUE into Sonata, we manage to export from Sonata P4-processed packets or datagrams filtered by Spark queries. In addition, since Sonata directly interfaces with P4 switches, we are also able to convert Sonata itself into an actuator of external P4 commands\cite{url:p4-cmd} that can be deployed on the switches it's monitoring. In companion to that, we develop a proof-of-concept firewall application that (1) consumes info of suspicious IP addresses (\eg SYN Flood attack\cite{url:syn-flood}) from Sonata queries, (2) matches it against \textbf{historical records} stored in SDNator's \textbf{Data Archives}, and (3) sends reconfiguration commands to Sonata to block attack traffic. See Workflow 1 in Figure~\ref{fig:workflows}.

Albeit an extremely simplified showcase of consuming both historical data and real-time data, the above use case can be profoundly extended by incorporating formal flow-based (offline using SDNator's Data Archives) or event-based (online using SDNator's Data Updates) methodologies described by Mooer \etal\cite{moore2006inferring}.

\subsubsection{Real-time Intrusion Detection and Mitigation}
\label{sec:kitsune}

Intrusion Detection Systems (IDS) have been popular targets\cite{liao2013intrusion} of the network security community and more recently of SDN\cite{sultana2019survey}.
For the second use case, we expand on an online, machine-learning-based network intrusion detection system (NIDS) called Kitsune\cite{mirsky2018kitsune} and implement a real-time intrusion detection and mitigation workflow (see Workflow 2 in Figure~\ref{fig:workflows}). We use P4 BMV2\cite{url:bmv2} software switches to emulate the data plane, and P4Runtime\cite{url:p4runtime} as the controller for the P4 switches. We stream packets from P4 switches to the P4Runtime controller and then pass over the packet stream to Kitsune using DUE; in return the controller receives analysis results from Kitsune via DUE and writes the results (either IP or MAC addresses to block) as flow table entries onto P4 switches.\looseness=-1

\begin{figure}[t!]
    \centering
    \includegraphics[width=0.99\columnwidth]{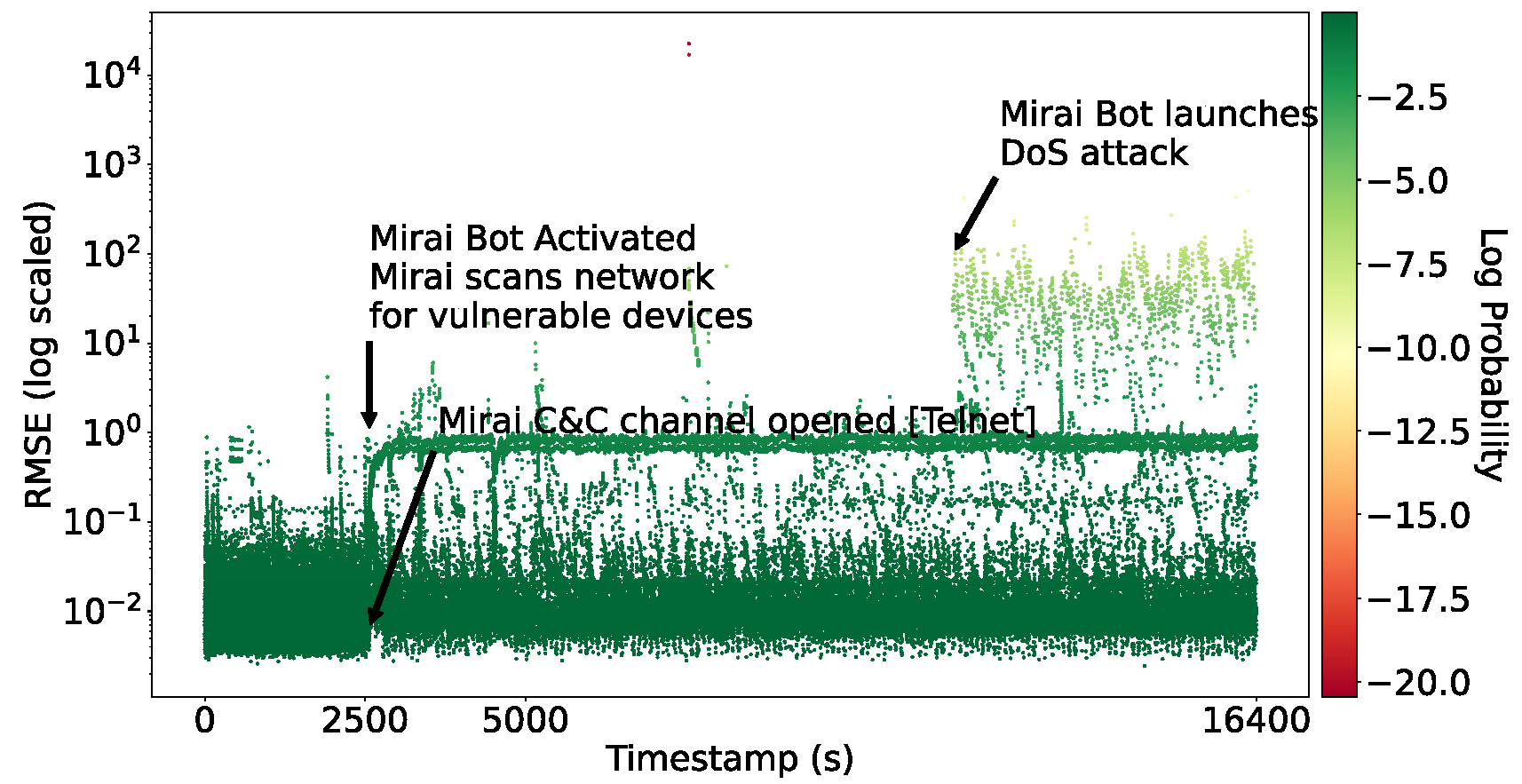}
    \vspace{-4mm}
    \caption{Using Kitsune as a standalone intrusion detection tool for Mirai botnet attacks.}
    \label{fig:kitsune-nonblocking}
\end{figure}

\begin{figure}[t!]
    \centering
    \includegraphics[width=0.97\columnwidth]{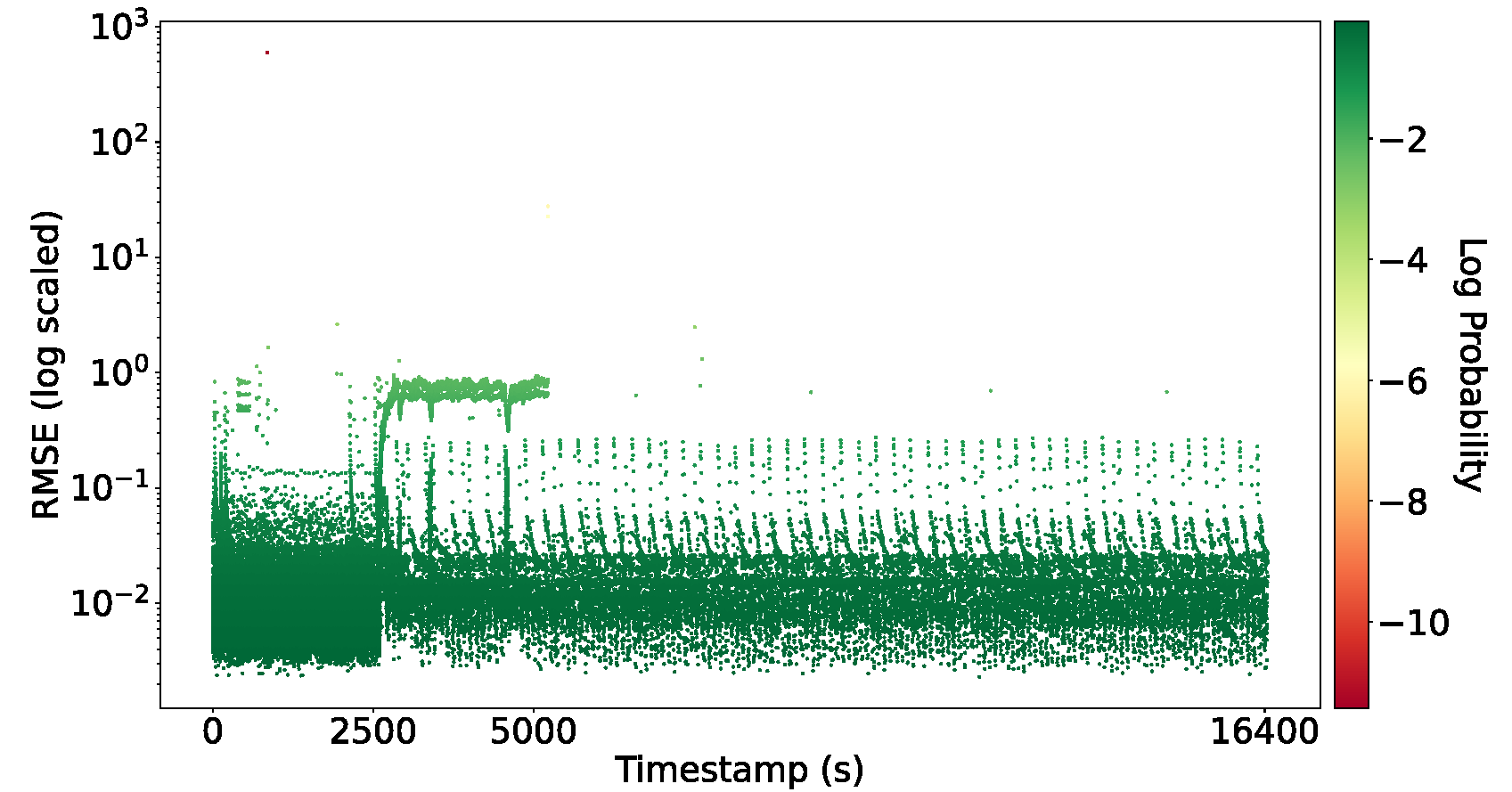}
    \vspace{-4mm}
    \caption{Using Kitsune in conjunction with P4Runtime to detect and block botnet attack traffic, enabled by SDNator.}
    \label{fig:kitsune-blocking}
\end{figure}

To verify the effectiveness of the implemented workflow, we test it with a simulated Mirai\cite{antonakakis2017understanding} botnet DoS attack using a dataset from Kitsune in an emulated network. By default, Kitsune uses the first \textasciitilde55k of packets in the stream to train its ML model, after which it will give an RMSE (anomaly) score to each packet. In Kitsune, an RMSE score of 1 or above can be considered an anomaly, so we modify Kitsune to publish source IP or MAC addresses of those packets scoring over 1. Subscribing to those results, P4Runtime instructs the P4 switches to drop packets from those addresses. Figures~\ref{fig:kitsune-nonblocking} and \ref{fig:kitsune-blocking} show time-series plots of the scores when Kitsune runs alone and as part of our workflow respectively. We can clearly see the difference made by the mitigation module from both figures. It is worth pointing out that we updated \textbf{<10 lines of code} in Kitsune to include it in our workflow using SDNator.

\vspace{-3mm}

\subsubsection{Takeaways}
There are a couple key observations we can make based on our networking case studies:
\begin{itemize}[topsep=2pt,itemsep=1pt,partopsep=2pt, parsep=2pt]
    \item Application integration with SDNator is usually straightforward with fewer than 10 lines of Python code.
    \item SDNator can easily interconnect heterogeneous modules/frameworks (\eg a P4 SDN controller and an IDS system) through its generic data-driven  pipelines and compose them into cooperative workflows.
    \item A common one-to-many data flow, where multiple apps consume the same packet stream, is naturally and efficiently supported by SDNator's publish-subscribe model.
    \item SDNator readily facilitates both online and offline data analysis use cases with Data Updates and Data Archives
    \item On-demand data production opens up unique dimensions for consumers to customize upstream data producers.
    \item Data-intensive workflows like real-time IDS can be supported with satisfying performance.
\end{itemize}



\section{Discussions}
\label{sec:disc}

\vspace{-2mm}

{\flushleft\textbf{Security and privacy concerns.}}
In SDNator, all applications communicate through DUE and the data backends. Without access control, an application can send data to and receive data from arbitrary applications, which could be exploited by malicious or compromised applications. Although security and privacy are not our main focus in this paper, and SDNator shares the same assumptions with existing SDN controllers that applications are run by trusted parties, we suggest a few changes that could increase the security of the system:
\begin{itemize}[topsep=2pt,itemsep=1pt,partopsep=2pt, parsep=2pt]
    \item The Coordinator maintains a per-key whitelist of allowed apps to restrict access to data keys.
    \item Set up a public authentication server that authorizes and manages access tokens for new apps.
    \item Add a proxy between data backends and DUE to enforce access control to the data backends by checking the access tokens as well as Coordinator's whitelist.
\end{itemize}

{\flushleft\textbf{Limitations.}}
As a distributed system, SDNator application throughput will be capped by available bandwidth. Since Data Updates and Archives can be scaled out, one could instantiate local data backends to facilitate applications in close proximity.

SDNator is also not an ideal tool for session-based communications where data flow is strictly point-to-point such as between clients and servers. Frameworks like gRPC\cite{url:grpc} are more suited for these scenarios.

Even though SDNator embraces data-driven use cases, it is not intended to be used to transfer data for the sake of transferring data, especially in bulk volumes. Traditional client-server solutions are more reliable with break-point resumes and caching.

\section{Conclusions}
\label{sec:conclusion}

\vspace{-2mm}

In this work, we go beyond building yet another ``SDN controller''. Instead, we provide researchers and developers an extensible, data-driven, scalable and easy-to-use platform to implement/integrate their applications and ``create'' their own centralized controllers for a cyber-physical system (CPS). SDNator supports both event-driven and data-driven programming patterns and allows apps to leverage both real-time data streams and historical data. Benchmarks show that SDNator delivers comparable performance to Ryu with minor overhead. We demonstrate SDNator's usability and generality through our case studies on networking and manufacturing CPS. Using SDNator, we carry out the first study on ditital-twin-based control of additive manufacturing fleets and see substantial performance gains in shortening job makespans in various scenarios.

To our knowledge, SDNator is the first open-source framework for CPS to enable easy development \& integration of apps for device management and data analysis in a centralized fashion. As the boundary between physical and traditional networking systems continues to blur (e.g., the increasing adoption of IoT devices), a platform that is domain-agnostic, highly scalable and easy to use will become even more relevant.\looseness=-1

\bibliographystyle{plain}
\clearpage
\bibliography{reference}

\end{document}